# A Sentence-Level Approach to Understanding Software Vulnerability Fixes

AMIAO GAO, Southern Methodist University, USA
ZENONG ZHANG, The University of Texas at Dallas, USA
SIMIN WANG, Southern Methodist University, USA
LIGUO HUANG, Southern Methodist University, USA
SHIYI WEI, The University of Texas at Dallas, USA
VINCENT NG, The University of Texas at Dallas, USA

Understanding software vulnerabilities and their resolutions is crucial for securing modern software systems. This study presents a novel traceability model that links a pair of sentences describing at least one of the three types of semantics (triggers, crash phenomenon and fix action) for a vulnerability in natural language (NL) vulnerability artifacts, to their corresponding pair of code statements. Different from the traditional traceability models, our tracing links between a pair of related NL sentences and a pair of code statements can recover the semantic relationship between code statements so that the specific role played by each code statement in a vulnerability can be automatically identified. Our end-to-end approach is implemented in two key steps: *VulnExtract* and *VulnTrace*. *VulnExtract* automatically extracts sentences describing triggers, crash phenomenon and/or fix action for a vulnerability using 37 discourse patterns derived from NL artifacts (CVE summary, bug reports and commit messages). *VulnTrace* employs pre-trained code search models to trace these sentences to the corresponding code statements. Our empirical study, based on 341 CVEs and their associated code snippets, demonstrates the effectiveness of our approach, with recall exceeding 90% in most cases for NL sentence extraction. *VulnTrace* achieves a Top5 accuracy of over 68.2% for mapping a pair of related NL sentences to the corresponding pair of code statements. The end-to-end combined *VulnExtract+VulnTrace* achieves a Top5 accuracy of 59.6% and 53.1% for mapping two pairs of NL sentences to code statements. These results highlight the potential of our method in automating vulnerability comprehension and reducing manual effort.

## 1 INTRODUCTION

Cyberattacks, including ransomware incidents, have increasingly targeted critical infrastructure and companies, leading to significant financial losses and public crises [55]. The exposure of millions of records due to security vulnerabilities highlights the urgent need for robust cybersecurity measures [59]. Understanding the causes of vulnerabilities and the logic behind their fixes is essential for improving software security, automating vulnerability fixes, and generating effective security patches [7]. Applications such as automated bug-fix generation and vulnerability patching rely heavily on the ability to comprehend the root causes and reasoning behind vulnerabilities, which, when understood well, can streamline the software maintenance process [27, 29, 36, 58].

The reasons for vulnerability occurrence and their fixes are often complicated, involving design- or implementation-specific issues that affect both the syntax and semantics of the program [58, 65]. The complexity of these issues is heightened by the fact that the relevant information is scattered across natural language (NL) artifacts, such as bug reports, Common Vulnerabilities and Exposures (CVE) [2] summaries, and commit messages, as well as code artifacts, such as commits or diffs that fix the bugs [18, 24, 43]. NL artifacts frequently explain the cause and crash phenomenon associated with vulnerabilities, while the code provides the actual fix [5, 43]. The challenge is in bridging these disparate sources to achieve a holistic understanding of how vulnerabilities are caused and fixed.

Authors' addresses: Amiao Gao, amiaog@smu.edu, Southern Methodist University, Dallas, Texas, USA; Zenong Zhang, The University of Texas at Dallas, Richardson, Texas, USA, zenong@utdallas.edu; Simin Wang, siminw@mail.smu.edu, Southern Methodist University, Dallas, Texas, USA; Liguo Huang, lghuang@smu.edu, Southern Methodist University, Dallas, Texas, USA; Shiyi Wei, The University of Texas at Dallas, Richardson, Texas, USA, swei@utdallas.edu; Vincent Ng, The University of Texas at Dallas, Richardson, Texas, USA, vince@hlt.utdallas.edu.

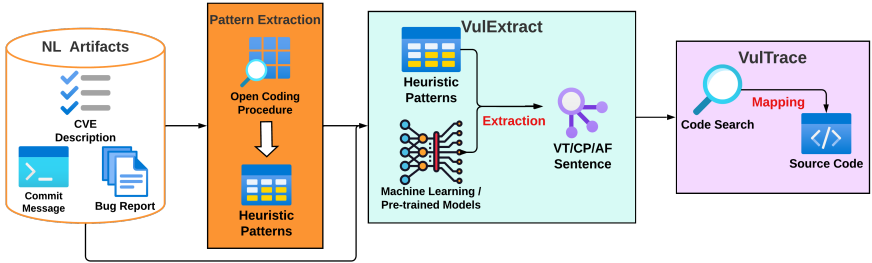

Fig. 1. Approach overview.

Existing efforts that address this problem tend to focus on either NL artifacts [5, 22, 39, 44, 64, 67] or code artifacts [7–9, 34, 40, 42, 49], but rarely on both in conjunction. Studies analyzing NL artifacts such as bug reports often highlight the importance of well-written bug reports for effective bug fixes [5], while others focus on automatically discovering vulnerabilities from textual descriptions in commit messages and CVE summaries [67]. These approaches fail to capture the underlying code logic that implements the fixes [27, 29, 43]. Empirical studies on code patterns have identified common bug-fix patterns and automatic program repair techniques that focus on code changes associated with vulnerabilities [7–9, 34, 40, 42, 49]. These approaches often overlook the rich contextual information present in NL artifacts, which can provide critical insights into the reasoning behind the fixes [5, 24, 33, 50]. Bug reports frequently contain crucial explanations of the symptoms of a vulnerability, the conditions under which it occurs, and the user-reported impacts, all of which can guide developers in identifying and prioritizing fixes [5]. Misclassifications, as discussed in Herzig et al.'s study [24], demonstrate how the correct interpretation of bug reports can dramatically influence the outcome of a bug fix, where certain issues that are classified as features instead of bugs delayed essential fixes. Similarly, studies have shown that bug report summaries help developers quickly locate relevant issues and understand the rationale behind code changes [33]. In addition, Rastkar et al. [50] found that summarizing bug reports effectively contributes to faster and more efficient bug resolution by filtering out extraneous information and providing developers with a concise overview of the critical issue.

To bridge this gap, we propose a novel traceability model (as shown in Fig. 1) that traces a pair of related NL vulnerability entities that describe a *v*ulnerability *t*rigger (VT), its *c*rash *p*henomenon (CP), and how it is repaired *a*fter being *f*ixed (AF) to their corresponding source lines of code. Our work presents the first end-to-end traceability approach to produce the relationships critical for understanding vulnerabilities and their fixes in both NL and code artifacts. We achieve this in two steps.

First, we develop *VulnExtract*, which leverages a catalog of 37 discourse patterns to automatically extract paired vulnerability entities – VT, CP, and AF – from NL artifacts such as bug reports, commit messages, and CVE summaries. These discourse patterns, designed through a qualitative analysis of vulnerability artifacts, act as linguistic structures that identify key elements describing the bug, its crash phenomenon, and its resolution. In addition, we integrate these patterns as features in machine learning models to enhance the precision of extracting these entities. These paired entities provide a detailed explanation of the trigger of the bug, the crash phenomenon, and the post-fix actions.

Second, we develop *VulnTrace*, which leverages pre-trained code search models to trace each of these entities back to their corresponding lines of code. This technique enables a comprehensive mapping between the NL artifacts and the specific code that implements the fix. The tracing link between a pair of related NL vulnerability entities and a pair of related lines of code automates the manual analysis process followed by a human developer to comprehend "why" and "how" a vulnerability bug is fixed.

We conducted extensive empirical experiments to evaluate the effectiveness of our approach, using a dataset of 342 labeled CVE records, which include corresponding bug reports and commit messages, amounting to 2,439 sentences. The dataset was annotated with our VT, CP, and AF labels. To evaluate the performance of *VulnExtract*, we applied both heuristic and machine learning-based extraction methods, incorporating our discourse patterns as features. The results demonstrated that our model achieves average recall levels of 100.00%, 95.01%, and 91.03% in accurately identifying and extracting VT, CP, and AF entities from NL artifacts respectively. For the *VulnTrace* step, we evaluated the accuracy of our code search models in tracing these extracted NL entities to their corresponding code lines. Our experiments demonstrated that VulnTrace achieves an accuracy of 68.20% for VT/CP_Code mapping and 68.33% for AF/CP_Code mapping, showcasing its effectiveness in tracing extracted NL entities to their corresponding code segments. The key takeaway from our experiments is that combining NL analysis with code search enhances traceability between vulnerability descriptions and their corresponding fixes, offering a more comprehensive understanding of how and why vulnerabilities are fixed. Our approach automates vulnerability comprehension and significantly reduces the manual effort for human analyzers to comprehend vulnerabilities. The main contributions of the paper are as follows.

- We have developed and generalized a catalog of 37 discourse patterns that can be used to extract sentences describing VT, CP, and/or AF from natural language vulnerability artifacts.
- We have conducted empirical experiments to identify the most effective approach (SVM + discourse patterns for VT, CP and AF) for automatically extracting the three types of sentences mentioned above.
- We propose a novel traceability model that can trace between a pair of related Natural Language (NL) vulnerability entities and their corresponding source lines of code.
- We have created a dataset of 342 labeled CVE records with corresponding bug reports and commit messages, with a total of 2,439 sentences identified as VT, CP and/or AF.
- A replication package including all data and complete results of the empirical experiments is provided with this study.

## 2 MOTIVATING EXAMPLE

Abby, a vulnerability analyst, needs to understand both why and how a vulnerability bug was fixed. Using a code diff tool, she quickly identifies where the vulnerability was fixed by examining the newly added lines of code (e.g., Line 19 marked with a leading + sign in Fig. 2b).

To fully grasp the reason behind the bug and its consequences, Abby seeks to explore the conditions that triggered the bug and where in the code it caused system crashes. Unfortunately, merely reviewing the code diff results or analyzing the source code itself (e.g., through syntactic or semantic analysis) cannot easily provide these answers. To gain a more comprehensive understanding, she turns to Natural Language (NL) artifacts such as CVE summaries, commit messages, and bug reports. In a traditional manual process, Abby attempts to locate both the bug's trigger and its crash locations within the source code, while also analyzing the commit message to understand the fix in detail. Fig. 2 illustrates this process using a real-world vulnerability, CVE-2017-12893 [1], showing the bug's fix through a commit message. (Fig. 2a) alongside the corresponding code snippet (highlighted in Fig. 2b). This process is labor-intensive, requiring Abby to switch back and forth between the commit message and the code to manually map each sentence from the commit message to the corresponding code lines. Such a manual mapping can take a significant amount of time, and deriving useful patterns, like the ones derived in existing works [66], would require her to analyze hundreds of similar vulnerability cases—an effort that could take months.

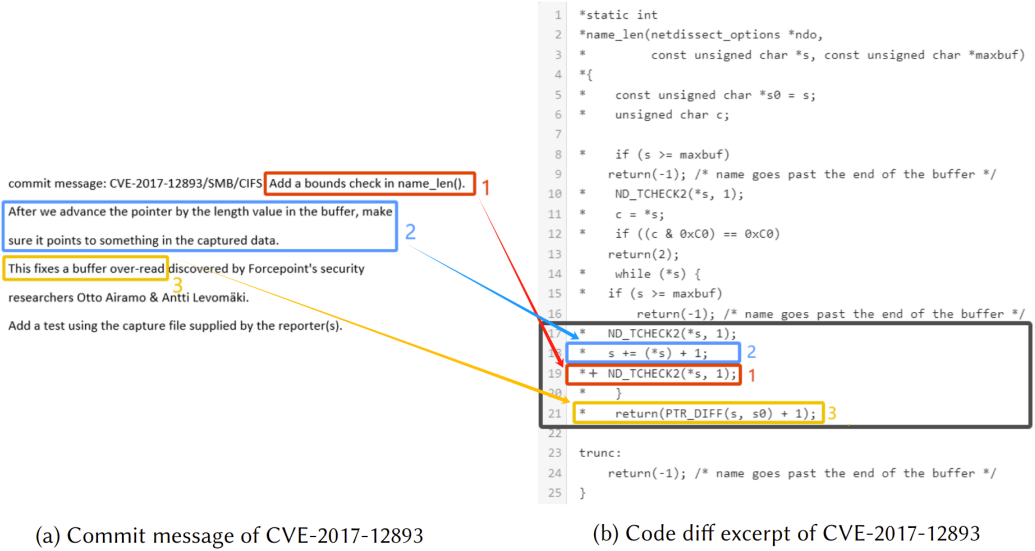

(a) Commit message of CVE-2017-12893    (b) Code diff excerpt of CVE-2017-12893

Fig. 2. Motivating example of CVE-2017-12893 [1].

Our approach aims to streamline this process by automating the analysis for Abby, helping her understand both why and how a vulnerability bug was fixed. It achieves this in two key steps:

First, the approach automatically extracts three types of vulnerability entities (**VT**, **AF**, and **CP**) from natural language artifacts (as described in Section 1). Specifically, **VT** refers to a sentence that reports the trigger (origin) of a vulnerability issue in the source code before the implementation of a bug fix, while **AF** refers to a sentence that explains how a vulnerability bug is fixed after the implementation of a bug fix. Lastly, **CP** refers to a sentence that describes a system crash phenomenon (e.g., Buffer Overflow, Null Pointer Dereference, Use After Free, Division by Zero, etc.) caused by a vulnerability in the source code. For instance, in the commit message from Fig. 2a) the sentence "*Add a bounds check in name_len().*" (in a red box) indicates the corrective action taken to fix the vulnerability, which corresponds to the **AF** entity that describes how the vulnerability was resolved. The sentence "*After we advance the pointer by the length value in the buffer, ensure it points to something within the captured data.*" (in a blue box) refers to the trigger of the bug, exemplifying the **VT** entity, which details the origin of the vulnerability in the source code. Finally, the sentence "*This fixes a buffer over-read...*" (in an orange box) describes the crash caused by the bug, representing the **CP** entity. These entities—**AF**, **VT**, and **CP**—are interconnected and pertain to the same vulnerability.

Second, the approach automatically traces each of these entities back to their corresponding code lines in Fig. 2b) as follows.

- **AF:** "*Add a bounds check in name_len().*" — `Line 19: ND_TCHECK2(*s, 1);`
- **VT:** "*After we advance the pointer by the length value in the buffer, ensure it points to something within the captured data.*" — `Line 18: s += (*s) + 1;`
- **CP:** "*This fixes a buffer over-read...*" — `Line 21: return(PTR_DIFF(s, s0) + 1);`

Since these entities are interrelated, the tracing links make it clear to Abby that Line 18, which advances the pointer ('`s += (*s) + 1;`'), triggered the vulnerability bug that led to a system crash at Line 21 ('`return(PTR_DIFF(s, s0) + 1);`'). This calculation of pointer differences prevents out-of-bounds memory access. Additionally, the newly added Line 19 ('`ND_TCHECK2(*s, 1);`'), which performs a bounds check, fixes the bug by ensuring the pointer remains valid after each advancement.

In essence, our approach automatically generates tracing links between pairs of related natural language vulnerability entities and their corresponding code line pairs, making it significantly easier for analysts like Abby to understand and track bug fixes efficiently.

## 3 RESEARCH QUESTIONS

To establish a foundation for automatically generating traceability links between natural language (NL) vulnerability artifacts and their corresponding code line pairs, our study investigates the following research questions. We structure our study into four key phases: (1) identifying and analyzing discourse patterns in vulnerability artifacts, (2) evaluating their impact on classification, (3) assessing their effectiveness in code traceability, and (4) evaluating end-to-end classification and traceability system performance.

**Phase 1: Identifying and Analyzing Discourse Patterns in NL Vulnerability Artifacts.**

- **RQ1.** What and how many discourse patterns for AF, VT, and CP can be identified from the NL artifacts? What are the most frequent patterns for each? How do the most frequent VT, AF, and CP patterns correlate with each other?

**Motivation**: NL artifacts, such as CVE, commit messages, and bug reports, exhibit recurring linguistic patterns that structure vulnerability-related information. Identifying these discourse patterns provides essential features that can enhance automated analysis and classification. Additionally, understanding correlations between VT, AF, and CP patterns offers insights into how vulnerabilities are discussed and resolved in practice.

**Relationship to other RQs**: The findings from **RQ1** serve as an empirical basis for **RQ2**, where we investigate whether these patterns improve machine learning-based classification. **RQ3** assesses a traceability model built on the basis of these patterns.

**Phase 2: Evaluating the Impact of Discourse Patterns in Vulnerability Classification.**

- **RQ2.** How do pre-trained models (with and without *discourse patterns*), SVM models (with and without *discourse patterns*), and a purely heuristic-based approach compare in classifying sentences describing VT, CP, and AF in *VulnExtract*?

**Motivation**: After identifying discourse patterns in **RQ1**, we assess their effectiveness in improving the classification of vulnerability-related sentences. This question explores whether incorporating discourse patterns enhances model performance compared to traditional machine learning and heuristic-based methods.

**Relationship to other RQs**: The effectiveness of discourse patterns in classification (**RQ2**) directly influences **RQ4**, where we examine how well these classifications contribute to an end-to-end traceability system. Additionally, the results inform **RQ3** by determining the quality of classified vulnerability-related sentences that serve as input for traceability models.

**Phase 3: Improving Code Traceability with *VulnTrace*.**

- **RQ3.** How effective are code search models in tracing vulnerability artifacts to relevant code using *VulnTrace*?

**Motivation**: Beyond classification, security analysts need to trace vulnerability descriptions to the corresponding code changes. **RQ3** evaluates the performance of different code search models in establishing these links, helping to determine which approaches best utilize structured textual representations for vulnerability traceability.

**Relationship to other RQs**: **RQ3** builds upon **RQ2**, as accurately classified vulnerability-related sentences are critical inputs for traceability models. Furthermore, its findings contribute to **RQ4** by assessing how well *VulnTrace* performs when integrated with *VulnExtract*.

**Phase 4: Assessing the End-to-End Effectiveness of *VulnExtract* + *VulnTrace*.**

- **RQ4.** How effective is *VuleExtract+VulnTrace* in combination?

**Motivation**: The final step in our investigation is to evaluate the complete system. By integrating *VulnExtract* (classification) and *VulnTrace* (traceability), we assess whether our approach improves efficiency and accuracy in tracking vulnerabilities from NL artifacts to source code.

**Relationship to other RQs**: **RQ4** synthesizes insights from all previous questions. The discourse patterns identified in **RQ1**, their impact on classification in **RQ2**, and the traceability performance in **RQ3** all contribute to evaluating the system's overall effectiveness. In addition, this RQ addresses the practical implications of our approach, testing whether the combined system improves efficiency and accuracy in vulnerability comprehension.

## 4 THE DISCOURSE OF VULNERABILITY DESCRIPTIONS

We aim to identify a comprehensive set of discourse patterns, referred to in this study as **discourse patterns**, which are commonly found in sentences within vulnerability-related NL artifacts. A discourse pattern is a generalized linguistic structure or schema that signifies semantic relationships between terms. These patterns serve as formalized rules that can be applied to the identification of key concepts and conceptual relationships in NL text [38]. By employing these patterns, we aim to extract meaningful insights from NL vulnerability artifacts, facilitating a deeper understanding of the cause and resolution of software vulnerabilities.

To meet this objective, we answer the following research question through a qualitative and quantitative analysis:

- **RQ1.** What and how many discourse patterns for AF, VT, and CP can be identified from the NL artifacts? What are the most frequent patterns for each? How do the most frequent VT, AF, and CP patterns correlate with each other?

Section 4.1 describes the methodology employed in the open coding procedure and presents a well-defined set of *discourse patterns* for VT, AF, and CP. These empirical findings serve as the foundation for our automated relational traceability approach. The insights gained from **RQ1** inform the evaluation of discourse pattern integration in *VulnExtract* (**RQ2**), the effectiveness of code search models in *VulnTrace* (**RQ3**), and the overall performance of the combined *VulnExtract* + *VulnTrace* approach (**RQ4**).

### 4.1 Open Coding Analysis

We follow the open coding procedure [10] to identify the *discourse patterns*, which are described next.

*4.1.1 Data Collection.* We selected the dataset collected by Zhang et al. [66] for the following reasons. First, it is extensively studied within the fuzzing literature and comprises well-documented CVEs. Second, the majority of CVE records in this dataset include direct links to the source code diffs of the bug fixes, which facilitate detailed analysis. The Common Vulnerabilities and Exposures (CVE) system is a standardized catalog of publicly known cybersecurity vulnerabilities and exposures. Each entry, often referred to as the "CVE Full Description," provides a comprehensive analysis of the vulnerability, including its technical nature, exploitation mechanisms, potential impacts, and remediation strategies. This detailed information is crucial for security professionals, software vendors, and organizations to understand and address vulnerabilities effectively, ensuring better protection against cybersecurity threats. In addition, most of these CVE records are reproducible and come with call stacks, providing invaluable insights into the origins of the bugs and the specifics of the fixes [66].

The original dataset consists of 693 CVEs from six open-source programs: Binutils [47], FFmpeg [57], libarchive [28], libxml2 [48], systemd [56], and Tcpdump [19]. We refined this dataset to include only CVEs with corresponding bug reports or commit messages, resulting in a final dataset of 341 CVEs from the aforementioned six open-source programs: 38 from Binutils, 116 from FFmpeg, 24 from libarchive, 48 from libxml2, 13 from systemd, and 102 from Tcpdump.

Using git diff [17], we generated code snippets for each CVE by comparing the parent and fix commits provided in [66]. We then split the CVEs, bug reports, and commit messages into individual sentences, producing 2,439 sentences. Each CVE was analyzed alongside its corresponding code snippet, and the sentences were classified as Vulnerability-Trigger (**VT**), After-Fix (**AF**), or Crash-Phenomenon (**CP**). To ensure clarity and consistency in annotation, we adopted a sentence-level labeling approach. A statistical analysis of our dataset revealed that multi-sentence cases were infrequent: only 3 cases (3.66%) for VT, 6 cases (0.77%) for AF, and 0 cases for CP. Given their low occurrence, we followed a single-sentence annotation approach to maintain uniformity in classification while ensuring alignment with our downstream classification models. Specifically, 82 sentences were labeled as **VT**, 780 as **AF**, and 808 as **CP**.

*4.1.2 Coding Procedure.* Two authors, both Computer Science Ph.D. students, performed the labeling and discourse pattern development tasks over a 31-day period. Each author dedicated more than 400 hours to the process, averaging approximately 13 hours per day. The annotation process consisted of two main phases: (1) sentence labeling for AF, VT, and CP and (2) discourse pattern development and categorization.

*Pilot Study for Consistency:* Before beginning the annotation of the 341 CVEs in the dataset, a pilot study was conducted to establish a shared understanding of the coding framework. This involved the following steps: 1. Randomly selecting a subset of 50 additional CVEs not included in the dataset. 2. Independently analyzing the CVE descriptions to identify sentences corresponding to AF, VT, and CP and inferring discourse patterns from these sentences. 3. Calculating Cohen's Kappa values for agreement on positive or negative labels for AF, VT, and CP. If the Kappa value did not reach the "almost perfect agreement" threshold (>0.8) as defined by Landis and Koch [30], disagreements were discussed and resolved to achieve consensus. 4. Repeating the above steps with a new set of 50 CVEs until the Kappa value exceeded 0.8. After two iterations, the Cohen's Kappa values for identifying AF, VT, and CP were 0.83, 0.84, and 0.84, respectively, which was deemed sufficient to begin annotating the full dataset.

The annotation process for the 341 CVEs involved: Labeling Sentences: Over 14 days, each author independently labeled all sentences in the dataset as positive or negative for AF, VT, and CP. Any disagreements were collaboratively reviewed and resolved to reach a consensus. Pattern Development: Over the remaining 17 days, the authors developed discourse patterns to categorize the positive sentences. A shared spreadsheet was maintained to document newly identified patterns. When one author identified a new pattern, it was recorded in the spreadsheet, and the other author was notified for verification. Independent Assignments and Resolution: The assignment of a discourse pattern to each positive sentence was done independently by both authors. Any disagreements in pattern assignment were subsequently reviewed and resolved through collaborative discussion.

To ensure the robustness and generalizability of our *discourse patterns*, we adopted a leave-one-out cross-validation (LOOCV) [4] pattern development approach across six open-source datasets: Binutils, FFmpeg, libarchive, libxml2, systemd, and Tcpdump. The core idea was to iteratively use one dataset as a testing set, while developing and refining patterns using the remaining five datasets as the training set. This procedure allowed us to focus on patterns that generalize well across different datasets and avoid overfitting to any specific one. **Data Partitioning**: The entire dataset consisting of 341 CVEs from six programs was divided into six subsets, each corresponding to one of the programs: Binutils (38 CVEs), FFmpeg (116 CVEs), libarchive (24 CVEs), libxml2 (48 CVEs), systemd (13 CVEs), and Tcpdump (102 CVEs). **Iteration Process**: In each iteration, one of the six datasets was excluded as the testing set, and the remaining five datasets were combined to form the training set. This process was repeated six times, with each dataset serving as the testing set once. **Pattern Learning and Refinement**: (1) **Training Phase**: For each iteration, discourse patterns were identified and refined based on the five training datasets. These patterns were developed by manually analyzing sentences that describe VT, AF, and/or CP. (2) **Testing Phase**: Once patterns were established using the training set, they were applied to the excluded testing set. Importantly, patterns or keywords that only appeared in the testing set and were not observed in any of the training datasets were excluded. This step ensured that patterns used for entity extraction were not overly dependent on any one dataset and were truly generalizable. **Exclusion of Testing-Specific Patterns**: Any patterns or keywords found only in the testing set and not present in the five training sets were excluded during the pattern application phase. This ensured that the discourse patterns were robust and not overfitted to any specific dataset.

To minimize subjectivity, both coders independently analyzed all 341 CVEs, marking each sentence as VT, AF, or CP if it indicated the description of VT, AF, or CP. They also map each VT, AF, and CP to their corresponding code snippet generated by git diff. Sentences could potentially fall into multiple categories simultaneously. *Discourse patterns* were then inferred from the marked sentences, and each pattern was assigned a unique code, uniquely identifying a *Discourse pattern*. It is possible to infer more than one pattern from a sentence. An online spreadsheet facilitated the sharing of pattern catalog between the two coders, enabling the rescue or addition of new patterns. When one coder independently identified a new pattern, it was included in the catalog and the other coder was notified. The newly identified patterns were verified by both coders. Similar patterns were consolidated into a more general pattern, and existing labels were updated accordingly, with the merging overseen by both of the coders. Any disagreements were addressed through collaborative discussions until a consensus was reached. Additionally, two domain experts in SE provided advice on "Hard to Determine" patterns. This iterative process involved continuous refinement of the pattern catalog and resolution of ambiguous cases.

In total, out of 42 initial patterns identified, 9 patterns were merged into 4 more general patterns, resulting in a total of 37 unique patterns — 6 for VT, 16 for AF, and 15 for CP.

*4.1.3 Coding Criteria and Agreement.* A list of coding criteria was provided for two coders and some important criteria were summarized below. Regarding VT, sentences that describe an issue or error caused by code that fails to perform a required action should be included here. This could involve code that is incomplete, lacks validation, or introduces potential bugs due to missing checks or faulty logic. Regarding AF, sentences that describe actions taken to modify or update code should be included under this category. These sentences often indicate changes to existing software functionality, such as adding, removing, or updating a feature, method, or check. Regarding CP, sentences that describe crash phenomena should be included here. These descriptions often focus on specific security issues like buffer overflows, null-pointer dereference, or use after free, etc. in the code. Inter-coder agreement was evaluated using Cohen's Kappa value. Our analysis reveals high levels of agreement, with Cohen's Kappa values exceeding 0.8 (0.83 for VT, 0.84 for AF, and 0.84 for CP), indicative of almost perfect agreement according to Landis and Koch.

*4.1.4 Coding Results and Analysis.* To address **RQ1**, the open coding approach resulted in a catalog of 37 discourse patterns, consisting of 6 VT patterns, 16 AF patterns, and 15 CP patterns, which are shown in [16]. Table 1 shows the most frequent AF, VT, and CP patterns. By comparing the AF, VT, and CP patterns, we examine the correlations below to address the second and the third part of **RQ1**.

Table 1. Top frequent discourse patterns for AF, VT, and CP.

| | AF | VT | CP |
|---|---|---|---|
| **Code** | AFBC | BFDN | CLBO |
| **Description** | Sentence describes applying a bound check to existing code. | Sentence describes code did not do a specific action(e.g. a proper variable initialization, a proper bound check, etc.) which would cause a bug. | Sentence indicates a buffer overflow in existing code. |
| **Rule** | [verb] [method/variable] [bound] [check] [adv] [method/variable] | [method/variable] [do not] [verb] [complement] | [buffer] [overflow] [adv] [method] |
| **Example** | CVE-2017-12897/ISO CLNS: Use ND_TTEST() for the bounds checks in isoclns_print(). | In lldp_private_8023_print() the case block for subtype 4 (Maximum Frame Size TLV, IEEE 802.3bc-2009 Section 79.3.4) did not include the length check. | The BGP parser in tcpdump before 4.9.2 has a buffer over-read in print-bgp.c:bgp_attr_print(). |

**VT patterns**. The three most frequent VT patterns account for 70.3 % of the CVEs that contain VT. In addition to the most frequent pattern (*BFDN*) shown in Table 1, the two other patterns are: (1) *BFWD*, which involves sentences using modal verbs like "will" or "would" to indicate that, under certain conditions, a vulnerability will occur, such as *"When 'uvalue' is a specific value, 'block_start + value' will cause integer overflow."*; and (2) *BFCU*, which corresponds to sentences using verbs like "remove" or "clean" to indicate that the removal of certain variables/statements/assignments can prevent the occurrence of a vulnerability, such as *"Remove use of FF_PROFILE_MPEG4_SIMPLE_STUDIO as an indicator of studio profile."*

**CP patterns**. The three most frequent CP patterns account for 72.9 % of the CVEs that contain CP. In addition to the most frequent pattern (*CLBO*) shown in Table 1, the two other patterns are: (1) *CLNP*, which involves sentences describing the crash phenomenon "null pointer dereference" to indicate the resolution of a crash, such as *"Fixes: null pointer dereference."*; and (2) *CLOOA*, which corresponds to sentences containing the phrase "OOB (out-of-bounds) access/read/write" to describe the mitigation of an OOB vulnerability, such as *"Check for size_t and vector resize() overflow to avoid OOB writes during vector allocation."*

**AF patterns**. The three most frequent AF patterns account for 34.6 % of the CVEs that contain AF. In addition to the most frequent pattern (*AFBC*) shown in Table 1, the two other patterns are: (1) *AFA*, which involves sentences using verbs like "avoid" or "reject" to describe the rejection or avoidance of certain variable assignments to fix a vulnerability, such as *"Reject vp8 video files that have alpha and image planes of different sizes."*; and (2) *AFR*, which corresponds to sentences using verbs like "adjust" or "set" to indicate a modification of a variable's assignment to fix a vulnerability, such as *"Set the default EXTINF duration to 1ms if the duration is smaller than 1ms."*

**Correlations**. Within vulnerability artifacts, we often observe patterns that frequently co-occur. The most frequent CP pattern, *CLBO*, as shown in Table 1, often appears alongside VT patterns that describe missing key actions, such as *BFDN*, and AF patterns that involve applying bounds checks, such as *AFBC* (both shown in Table 1). In these cases, the absence of proper validation, represented by *BFDN*, leads to unsafe memory access, which results in buffer overflows or over-reads, as captured by *CLBO*.

Furthermore, AF patterns like *AFBC*, which involve adding bounds checks (e.g., *"Use ND_TTEST() for the bounds checks in isoclns_print()"*), often co-occur as a corrective action following the identification of vulnerabilities where these validations were missing. This interplay between *BFDN*, *CLBO*, and *AFBC* highlights a recurring sequence in software vulnerabilities: the failure to implement appropriate bounds checks (*BFDN*) leads to memory-related errors (*CLBO*), which are then mitigated through corrective measures like bounds checks (*AFBC*).

This correlation illustrates how VT, CP, and AF patterns interact, with missed validation steps causing crashes and subsequent fixes involving bounds checks. Understanding these patterns underscores the importance of identifying and addressing VT issues early to prevent CP-related consequences, while also emphasizing the role of AF patterns in mitigating the vulnerabilities caused by such oversights.

## 5 APPROACHES

Our approach enables automated comprehension of *why* and *how* a vulnerability is fixed via two key steps: *VulnExtract* and *VulnTrace* as shown in Fig. 1.

### 5.1 VulnExtract: Automatically Extract Essential Information from Natural Language (NL) Vulnerability Artifacts

Manually extracting essential information describing VT/AF/CP from NL vulnerability artifacts is extremely effort-consuming. To meet this challenge, we aim to automate the extraction process. We propose and compare four approaches to automatically extract VT/AF/CP from the NL vulnerability artifacts. We will extract each sentence that contains VT, AF or CP (defined in Section 2). We choose a sentence as the level for linguistic analysis because it contains richer contextual information to describe each of the three types of entities. Note that sometimes a sentence may contain more than one type of entity.

*5.1.1 Heuristic Pattern Approach.* This approach adopts a heuristic pattern methodology wherein we implement the 37 manually identified discourse patterns to automatically apply them to new CVE issues. Specifically, we assess whether any of the patterns are applicable for each sentence in a given CVE summary, bug report, and commit message. If a pattern applies to a sentence, it indicates that the sentence contains a description of VT, CP or AF. Consequently, the discourse pattern approach labels each sentence as containing VT, CP, or AF if and only if at least one pattern is applicable.

We employ heuristics to align sentences with our discourse patterns to implement the patterns. Each pattern from our catalog was implemented using the Natural Language Toolkit (NLTK) [6] in Python 3, utilizing the Jupyter Notebook platform for development and testing.

*5.1.2 Traditional Machine Learning.* Following the methodology outlined by Chaparro et al. [10], we experiment with a machine learning approach using Support Vector Machines (SVM). We implement our binary SVM classifier using the scikit-learn machine learning package [45]. In this approach, each sentence is treated as a training instance, with positive instances (sentences) labeled as VT, CP or AF, and sentences with no description of VT, CP or AF labeled as negative instances.

Two feature types were used to train the binary SVM classifier:

- **Discourse patterns** These are pattern features derived from the heuristic approach detailed in Section 4.1. Each VT, CP and AF pattern is represented as a boolean feature, indicating whether the sentence contains a description of VT, CP or AF, respectively. VT, CP and AF features are employed by the VT classifier, CP classifier and AF classifier, respectively.
- **N-grams** These features represent contiguous sequences of n terms within a sentence. We incorporated unigrams, bigrams, and trigrams, transforming each sentence into a vector of counts, indicating its presence or absence within the sentence.

We optimize the hyperparameters of the SVM classifier using grid search, as implemented in the scikit-learn [45] package's GridSearchCV function. The hyperparameter grid explored includes the following settings. We tune the penalty parameter C of the SVM classifiers via grid search on the parameter tuning fold. We experiment with 20 evenly spaced parameter values over the interval [0.001,100]. The optimal C is selected by maximizing the F1 score of the trained SVMs to identify VT, CP, and AF entities in vulnerability artifacts.

*5.1.3 Pre-trained Models.* Pre-trained models, which are trained on self-supervised learning tasks using a large amount of unlabeled data, possess a lot of linguistic and commonsense knowledge and have been shown to outperform traditional machine learning approaches, including early deep learning approaches (e.g., RNNs such as LSTMs), on a variety of NLP and SE tasks [15, 23, 32]. Hence, we choose pre-trained models as one of our candidate approaches to indicate if the sentence contains a description of VT, CP or AF.

Specifically, we chose five state-of-the-art pre-trained models from Hugging Face library [62], namely BERT [13], DistilBERT [14], ELECTRA [25], SpanBERT [26], RoBERTa [35], and fine-tuned them on the training portion of our data. Each training sentence corresponds to a CVE, labeled as positive if and only if the sentence is identified as VT, CP, or AF. Each sentence is represented as a sequence of words in the corresponding VT, CP or AF description. In other words, manual feature engineering is not needed for pre-trained models. Nevertheless, recent work on SE tasks has shown that augmenting the input word sequence with hand-crafted features can often yield improved performance [31, 60]. Hence, we additionally experiment with a version of our pre-trained models where each of them will be provided with hand-crafted features.

The question, then, is: what hand-crafted features should be used? Given that the SVM experiments employed two types of features, *discourse patterns*, and n-grams, and that n-grams can be readily extracted from the input word sequence, it is a natural choice to use the *discourse patterns* as features for the pre-trained models. We derive features from the *discourse patterns* for the pre-trained models in the same way as in the SVM experiments, meaning that we employ one binary feature for each pattern whose value is 1 if and only if the pattern is applicable to the sentence to which the training instance corresponds to. The resulting binary-valued feature vector is then concatenated with the vector encoding the input sentence, and the resulting vector is then used for extracting VT, CP and AF. We used the 12-layer and 768-hidden version for all models and tuned the dropout rate for each pre-trained model using the parameter tuning set, considering the following parameter values: {0.1, 0.2, 0.3, 0.4, 0.5}.

## 5.2 VulnTrace: A Novel Traceability Model between NL Vulnerability Artifacts and Code

Traditional traceability models trace between a single NL entity (e.g., requirement, bug report) and code. Therefore, they lack the ability to recognize each line of code's specific role when a pair of related lines of code need to be comprehended in vulnerability analysis, such as identifying which lines of code contain the bug and which ones show the crash location [3]. This limits their ability to explain the underlying "Why" and "How" of a software vulnerability fix.

To overcome this challenge, we propose a novel traceability model, *VulnTrace*, which not only establishes a tracing link between an NL vulnerability entity (i.e., VT, CP or AF) and the corresponding code lines but also understands the specific role each line of code plays. Unlike traditional traceability models that only trace between a single entity to code or between a pair of code segments, our new traceability model traces pairs of related vulnerability entities to corresponding pairs of code lines. At the same time, the semantic relation between the pair of vulnerability entities are maintained. We cast this traceability problem into a code search task described in Section 4.2.1. Specifically, a sentence labeled as VT, CP or AF by *VulnExtrat* is used as the search query for the code search task, and the code snippets generated by Git Diff are used as a pool of code candidates where to search from. Finally, the code search results return the corresponding lines of code that implement VT, CP or AF. In this way, a tracing link between a pair of related vulnerability entities (VT/CP, CP/AF or VT/AF) and a pair of code lines that implement each of these entities is established. This allows us to identify the specific role (i.e., vulnerability trigger, crash phenomenon, or fix action) each line of code takes and understand the rationales behind a software vulnerability.

*5.2.1 Code Search.* Pre-trained models show their robust abilities in code search tasks. Hence, we choose pre-trained code search models in *VulnTrace*. Specifically, we chose six state-of-the-art pre-trained models from the Hugging Face library [62], namely UniXCoder [20], CoCoSoDa [53], GraphCodeBERT [21], CodeBERT [15], RoBERTa [35], RoBERTa-Code [37].

To maintain the fairness of the experiment, we follow the configuration guidelines and apply identical parameter settings across all models. Choosing a base pre-trained model offers a balanced trade-off between performance and computational efficiency, making it suitable for a wide range of real-world applications. Base models require less computational power and memory, leading to faster training and inference times, which is beneficial for resource-constrained environments and cost-effective deployment. Thus, we use all the base versions for each pre-trained model. We fine-tuned all six pre-trained models to optimize their effectiveness in the *VulnTrace* task. Each model was trained using a learning rate of 2e-5, with AdamW as the optimizer, a batch size of 32, and 10 epochs. Following the LOOCV approach, during each iteration, the model was trained on five datasets while reserving one dataset for testing.

## 6 EXPERIMENTAL DESIGN

To evaluate the effectiveness of *VulnExtract* in identifying VT, CP and AF from NL vulnerability artifacts, and *VulnTrace* in mapping these entities to the corresponding code segments, we have formulated three research questions:

- **RQ2.** How do pre-trained models (with and without *discourse patterns*), SVM models (with and without *discourse patterns*), and a purely heuristic-based approach compare in classifying sentences describing VT, CP, and AF in *VulnExtract*?
- **RQ3.** How effective are code search models in tracing vulnerability artifacts to relevant code using *VulnTrace*?
- **RQ4.** How effective is *VuleExtract+VulnTrace* in combination?

For **RQ2**, we employ LOOCV to conduct 6 sets of experiments,

- **Set 1.** Only apply heuristic approach
- **Set 2.** Apply the SVM approach with the inclusion of n-grams as features
- **Set 3.** Apply the SVM approach with the inclusion of discourse patterns as features
- **Set 4.** Apply the SVM approach with the inclusion of both n-gram and discourse patterns as features
- **Set 5.** Only apply the pre-trained model approach
- **Set 6.** Apply the pre-trained model approach with the inclusion of discourse patterns as features

For **RQ3** and **RQ4**, we apply LOOCV to train all pre-trained code search models. For **RQ3**, we use *VulnExtract*'s ground truth, and for **RQ4**, we use the VT, CP and AF sentences extracted by *VulnExtract* as search queries, with the code snippets output by Git Diff serving as potential code mapping candidates.

**Evaluation Metrics.** We measure *precision*, *recall*, and *F1 score* to evaluate the effectiveness of *VulnExtract* in identifying sentences describing VT, CP or AF from NL vulnerability artifacts. *Precision* is defined as the percentage of sentences predicted as VT/CP/AF that are correct according to the gold set. *Recall* measures the percentage of VT/CP/AF sentences that are correctly predicted.

The *F1 score*, which is the harmonic mean of precision and recall, provides a combined measure of accuracy.

We use the *TopK* metric to evaluate the accuracy with which each VT, CP or AF sentence can be mapped to code lines appearing within the top *K* results returned by each *VulnTrace* model. Considering VT, CP, and AF sentences are extracted from multiple NL vulnerability artifacts so that multiple sentences could convey the same semantics and be mapped to the same code lines, we claim the code search by such group of sentences with equivalent semantics a *hit* if any sentence in the group is correctly mapped to any one of the code lines that implement the semantics.

Recall that the ultimate goal of our novel traceability model is to trace between a pair of related vulnerability entities extracted from NL artifacts to a pair of related code lines. Thus, we also measure the *TopK* accuracy of mappings between such relation pairs. Specifically, we compute the *TopK* for the AF/CP_Code and VT/CP_Code mappings[1] generated by *VulnTrace*. Here AF/CP_Code denotes a mapping between a pair of related AF and CP sentences that describe the fix action and crash phenomenon of the same vulnerability to a pair of code lines implementing the corresponding semantics. Similarly, VT/CP_Code denotes a mapping between a pair of related VT and CP sentences that describe the trigger and crash phenomenon of the same vulnerability to a pair of code lines implementing the corresponding semantics. We introduce two variants of the *TopK* metric, assuming a perfect and imperfect *VulnExtract*, respectively. *Formula (1)* evaluates *VulnTrace* on the gold set (**RQ3**) while *formula (2)* applies to *VuleExtract+VulnTrace* where the false positive results returned by *VuleExtract* is taken into account (**RQ4**):

$$(1) \quad \frac{\sum_{i=1}^{N} \mathbf{a}_i^T \cdot \mathbf{cp}_i^T}{\sum_{i=1}^{N} \mathbf{A}_i^T \cdot \mathbf{CP}_i^T} \qquad (2) \quad \frac{\sum_{i=1}^{N} \mathbf{a}_i \cdot \mathbf{cp}_i}{\sum_{i=1}^{N} (\mathbf{A}_i^T + \mathbf{A}_i^F) \cdot (\mathbf{CP}_i^T + \mathbf{CP}_i^F)}$$

where:

- **N**: the total number of CVEs.
- $\mathbf{a}_i$: the number of correct mappings between relevant AF or VT sentences (true positives) identified by **VulnExtract** and the code statements for the *i*-th CVE.
- $\mathbf{cp}_i$: the number of correct mappings between relevant CP sentences (true positives) identified by **VulnExtract** and the code statements for the *i*-th CVE.
- $\mathbf{a}_i^T$: the number of correct mappings between all AF or VT sentences in the gold set and the code statements for the *i*-th CVE.
- $\mathbf{cp}_i^T$: the number of correct mappings between all relevant CP sentences in the gold set and the code statements for the *i*-th CVE.
- $\mathbf{A}_i^T$: the total number of mappings between AF or VT sentences and code statements in the gold set for the *i*-th CVE.
- $\mathbf{A}_i^F$: the number of AF or VT sentences incorrectly identified by **VulnExtract** (false positives) for the *i*-th CVE.
- $\mathbf{CP}_i^T$: the total number of mappings between CP descriptions and code statements in the gold set for the *i*-th CVE.
- $\mathbf{CP}_i^F$: the number of CP sentences incorrectly identified by **VulnExtract** (false positives) for the *i*-th CVE.

---

[1] In our performance evaluation, we focus exclusively on AF/CP_Code and VT/CP_Code, excluding AF/VT_Code. This is because VT_Code mappings typically correspond to code deletions in a previous file version, while AF_Code mappings generally represent code additions in the newer version. Consequently, using both VT and AF code lines as entry points for static analysis (e.g., data dependency analysis) is not feasible, as the entry points must be present in the same version of the code. Evaluating *VulnTrace*'s performance on AF/VT_Code would, therefore, not yield meaningful insights and is omitted from our analysis.

Table 2. Average VulnExtract performance for AF, VT, and CP.

| Approach | Features | AF | | | VT | | | CP | | |
|---|---|---|---|---|---|---|---|---|---|---|
| | | avg. P. | avg. R. | avg. F1 | avg. P. | avg. R. | avg. F1 | avg. P. | avg. R. | avg. F1 |
| Heuristic | patterns | 84.16% | 87.51% | 85.67% | 34.61% | **100.00%** | 46.68% | 92.19% | 92.34% | 92.18% |
| SVM | n-gram | 84.78% | 55.65% | 66.73% | 0.00% | 0.00% | 0.00% | 86.58% | 67.02% | 74.65% |
| | patterns | **94.84%** | **91.03%** | **92.80%** | 69.32% | 40.20% | **47.37%** | **96.48%** | **95.01%** | **95.60%** |
| | n-gram+patterns | 94.56% | 89.95% | 92.16% | 66.67% | 40.20% | 47.08% | 96.36% | 90.92% | 93.42% |
| BERT | - | 75.44% | 78.69% | 77.07% | 15.00% | 7.06% | 7.43% | 85.75% | 82.08% | 83.51% |
| | patterns | 79.97% | 84.12% | 81.56% | 49.40% | 28.36% | 28.02% | 86.51% | 87.75% | 86.89% |
| DistilBERT | - | 77.85% | 84.68% | 80.52% | 17.78% | 8.18% | 9.01% | 84.28% | 86.99% | 84.78% |
| | patterns | 78.83% | 87.05% | 82.29% | 31.13% | 13.66% | 18.23% | 84.97% | 85.88% | 85.00% |
| ELECTRA | - | 77.62% | 86.50% | 81.72% | 19.45% | 9.29% | 10.63% | 84.79% | 84.09% | 84.00% |
| | patterns | 81.34% | 82.24% | 81.46% | 46.27% | 23.70% | 31.92% | 89.83% | 80.34% | 84.21% |
| SpanBERT | - | 75.98% | 88.21% | 81.56% | 19.45% | 7.46% | 8.89% | 85.66% | 84.67% | 84.84% |
| | patterns | 79.22% | 86.60% | 82.53% | 42.55% | 23.19% | 20.44% | 85.69% | 90.66% | 87.83% |
| RoBERTa | - | 79.42% | 85.36% | 82.14% | 17.22% | 8.18% | 8.49% | 85.32% | 85.27% | 84.74% |
| | patterns | 75.77% | 88.38% | 81.46% | 52.32% | 29.95% | 33.35% | 86.68% | 90.49% | 88.01% |

In *formula (1)*, the numerator counts the total number of correctly identified relationship mappings for AF/CP_Code or VT/CP_Code in all CVEs. The denominator counts the total number of relationship mappings for AF/CP_Code or VT/CP_Code in all CVEs in the gold set. The key difference of numerators between *formula (1)* and *(2)* lie in the input source of VT, CP and AF sentences to *VulnTrace*. Formula (1) applies when *VulnTrace* takes the input of VT, CP and AF sentences from the gold set. *Formula (2)* applies when *VulnTrace* takes the input of VT, CP and AF sentences correctly identified by *VulnExtract*. Additionally, in *formula (2)*, the denominator counts not only all the relationship mappings in the gold set but also AF and VT sentences incorrectly identified for each CVE (false positives) by *VulnExtract*.

## 7 EVALUATION

### 7.1 Results of Heuristics, SVM, and Pre-trained Models in VulnExtract (RQ2)

Table 2 presents the average results of three different approaches in VulnExtract: the heuristic approach (row 2), SVM-based models (rows 3-5), and pre-trained models (rows 6-15). The rows labeled "patterns" indicate that discourse patterns were included as features in those models.

Overall, **SVM + patterns** (row 4) achieved the highest F1-scores for AF and CP, reaching 92.8% and 95.6%, respectively. However, all models struggled with VT classification, with F1-scores remaining below 50%. The **heuristic approach** achieved perfect recall (100%) for VT, similar to SVM + patterns, but with lower precision. Below, we analyze each approach in detail.

*7.1.1 Heuristic Approach.* The heuristic approach demonstrated strong performance for AF (85.67% F1) and CP (92.18% F1), but its effectiveness in VT classification was lower, achieving an F1-score of only 46.68%. The key advantage of the heuristic approach is its high recall for VT (100%), ensuring that all VT instances are retrieved. High recall is particularly useful in traceability models, where missing relevant links could negatively impact tasks such as bug-fix identification and vulnerability management [12, 51].

The lower precision for VT can be attributed to the relatively small number of VT instances (82) compared to AF (780) and CP (808). Vulnerability artifacts tend to focus more on bug fixes and crash descriptions rather than the root causes of vulnerabilities [63], making VT harder to identify. To ensure broad applicability, the heuristic pattern for VT was designed in a generalized manner. This generalization explains why recall reaches 100%, as the broader patterns capture most VT instances, though sometimes at the cost of specificity.

To gain further insights, we analyzed false positives in VT classification. A prominent category of false positives includes sentences beginning with *"Fix."* These typically describe patches, corrective actions, or preventive measures rather than the root cause of a vulnerability. For example, in **Binutils**,

*"Fix incorrect escape sequence handling."* states that a fix was applied but does not explain how the issue arose. Similarly, in **FFmpeg**, *"Fix small picture upscale."* refers to a surface-level correction rather than the vulnerability trigger. These misclassifications occur because heuristic patterns capture corrective actions but do not distinguish between patches and root cause explanations. While fixing an issue is part of remediation, VT typically describes how a vulnerability emerged, what faulty logic allowed it, or the preconditions that led to its exploitability. Sentences stating only that something was *"fixed"* lack this context, leading to incorrect VT classification. Beyond sentences beginning with *"Fix."*, we identified other false positives where the heuristic approach misclassified code maintenance actions or speculative system behaviors as VT. For example, in **FFmpeg**, *"Clear pointers in allocate_buffers()."* describes a routine cleanup without indicating how a flaw was introduced or exploited. Similarly, in **libarchive**, *"Fuzzing with CRCs disabled revealed that a call to get_uncompressed_data() would sometimes fail."* was misclassified due to its conditional phrasing, despite lacking details on the vulnerability's origin. These cases highlight the challenge of distinguishing between root cause indicators and broader system behaviors. While certain keywords suggest vulnerability relevance, correctly identifying VT requires recognizing whether a sentence explicitly describes how a vulnerability arises, rather than just its symptoms or remediation.

*7.1.2 SVMs.* Rows 3-5 in Table 2 show the results for three **SVM-based models** that use different feature sets: n-gram features only, pattern-based features only, and a combination of both. The **SVM + patterns model** (row 4) achieved the highest F1 scores for AF (92.8%) and CP (95.6%), confirming that incorporating discourse patterns significantly enhances classification performance. However, VT classification remained a challenge, with an F1-score of 47.37%.

Notably, **SVM + n-grams** alone (row 3) failed to retrieve any correct VT instances, leading to an F1-score of 0%. This is due to the imbalance between true positive and true negative VT instances, where the scarcity of positive VT samples leads to biased learning. Compared to the heuristic approach, **SVM + patterns** achieved higher precision for AF and CP but traded off recall, highlighting the precision-recall trade-off between heuristics and machine learning models.

To further understand SVM model limitations, we examined false positives and false negatives. The false positives revealed that SVM + patterns suffered from the same misclassification issues observed in the heuristic approach, reinforcing the idea that VT identification is challenging due to insufficient information in vulnerability artifacts. False negatives were more varied across projects. In **Binutils**, *"When 'uvalue' is a specific value, 'block_start + uvalue' will cause integer overflow."* was overlooked, likely due to its numerical nature, which lacks explicit vulnerability terms. In **FFmpeg**, *"Failure to verify causes total_size > atom.size which will result in negative size calculations later on."* was missed as it describes an eventual failure rather than explicitly linking to a vulnerability. Similarly, in **Tcpdump**, *"The bounds check in esis_print() tested one pointer at the beginning of a loop that incremented another."* was misclassified, suggesting difficulty in detecting vulnerabilities arising from complex interactions rather than isolated operations. These cases highlight the challenge of identifying VT instances that rely on implicit descriptions or technical conditions.

*7.1.3 Pre-trained Models.* Rows 6-15 in Table 2 present the results for six pre-trained models. A few points deserve mention. First, comparing the results of the pre-trained models with the SVM results when discourse patterns are *not* used (i.e., for the SVMs, only n-grams are used as input, and for the pre-trained models, only the input word sequence is used), we can see that all of the pre-trained models substantially outperform the SVMs. This is perhaps not surprising: unlike the SVMs, where the one-hot feature representations capture only lexical knowledge, the distributed representations used by the pre-trained models capture semantic knowledge and provide better generalizations than their one-hot counterparts.

Table 3. Average VulnTrace only experimental results for VT_Code, AF_Code and CP_Code.

| Approach | avg. VT_Code % | | | | | avg. AF_Code % | | | | | avg. CP_Code % | | | | |
|---|---|---|---|---|---|---|---|---|---|---|---|---|---|---|---|
| | top1 | top2 | top3 | top4 | top5 | top1 | top2 | top3 | top4 | top5 | top1 | top2 | top3 | top4 | top5 |
| UniXCoder | **11.63** | 16.74 | 33.64 | 51.87 | **72.57** | **17.61** | **34.86** | **49.15** | **63.34** | **77.62** | 9.31 | 25.86 | **48.28** | **64.02** | **81.72** |
| CoCoSoDa | **11.63** | 14.55 | 22.90 | 34.71 | 48.51 | 9.12 | 20.06 | 31.73 | 42.70 | 55.16 | 8.30 | 22.01 | 30.81 | 42.88 | 56.15 |
| GraphCodeBERT | 6.95 | **18.73** | 33.27 | 39.34 | 66.23 | 3.68 | 10.36 | 25.45 | 39.19 | 51.86 | 9.12 | 18.51 | 39.83 | 57.03 | 68.73 |
| CodeBERT | 4.27 | 13.61 | **34.11** | **57.81** | 66.28 | 5.81 | 13.51 | 26.73 | 43.67 | 59.69 | 14.51 | 26.40 | 40.47 | 55.16 | 74.36 |
| RoBERTa | 4.18 | 10.93 | 31.79 | 52.25 | 60.85 | 6.77 | 16.47 | 29.87 | 42.52 | 61.46 | 11.82 | 24.70 | 39.08 | 56.02 | 72.28 |
| RoBERTa-Code | 10.53 | 16.47 | 25.02 | 41.62 | 69.65 | 6.11 | 13.22 | 27.29 | 40.95 | 58.98 | **16.73** | **27.76** | 42.19 | 55.63 | 73.05 |

Table 4. Average VulnTrace only experimental results for VT/CP_Code and AF/CP_ Code.

| Approach | avg. VT/CP_Code | | | | | avg. AF/CP_Code | | | | |
|---|---|---|---|---|---|---|---|---|---|---|
| | top1 | top2 | top3 | top4 | top5 | top1 | top2 | top3 | top4 | top5 |
| UniXCoder | **4.77%** | 8.81% | 23.19% | 35.10% | **68.20%** | **2.43%** | **12.22%** | **26.19%** | **42.02%** | **68.33%** |
| CoCoSoDa | 2.39% | 7.05% | 16.96% | 30.26% | 47.16% | 1.36% | 6.14% | 13.50% | 26.97% | 41.92% |
| GraphCodeBERT | 0.88% | **10.85%** | **30.02%** | 34.32% | 65.94% | 0.65% | 3.54% | 16.40% | 28.12% | 41.38% |
| CodeBERT | 0.00% | 5.21% | 18.67% | **43.93%** | 60.71% | 1.58% | 5.01% | 16.93% | 34.59% | 52.32% |
| RoBERTa | 2.07% | 4.16% | 21.68% | 34.94% | 50.04% | 1.43% | 9.27% | 24.20% | 33.77% | 50.11% |
| RoBERTa-Code | 2.51% | 5.46% | 10.66% | 33.08% | 65.12% | 2.23% | 4.92% | 12.60% | 30.07% | 49.56% |

Second, when discourse patterns are added to the pre-trained models, we see an increase in average F1-scores in a majority of cases. Among these cases, we see increases in recall accompanied by smaller drops in precision. This is perhaps not surprising either: the discourse patterns provide further generalizations beyond what can be learned from the training data, thus helping to boost recall. For the remaining cases, the addition of discourse patterns causes the pre-trained models to perform either at the same level or marginally worse, and these are typically the cases when the discourse patterns fail to improve recall.

Finally, comparing the results of the pre-trained models and those of the SVMs when discourse patterns are added, we see that the SVMs substantially outperform the pre-trained models. In other words, the incorporation of the discourse patterns offers substantial improvements to the SVMs, but their impact on the pre-trained models is only modest at best. The reason is that the n-grams are less sparse and less correlated with the class labels, thus making it easy for the SVMs to identify them as not being useful (in the presence of the discourse patterns), allowing the SVMs to rely mostly on the discourse patterns when making classification decisions. In comparison to the n-grams, the dense word vectors used by the pre-trained models are not sparse and are more correlated with the class labels. This makes it hard for the pre-trained models to focus largely on the discourse patterns and ignore the word vectors when making decisions. In other words, the reliance on the word vectors has moderated the (positive) effect of the discourse patterns in the pre-trained models.

### 7.2 Performance of code search models in VulnTrace (RQ3)

Table 3 provides a detailed comparison of different code search models (UniXCoder, CoCoSoDa, GraphCodeBERT, CodeBERT, RoBERTa, and RoBERTa-Code) based on their average performance among all programs in mapping single sentence (VT, AF, CP) to corresponding code segments. The models are evaluated across different performance levels (top1 - top5). We find that **UniXCoder** outperforms the other models across most of the entity types in Table 3, particularly due to its cross-modal pre-training that efficiently handles both natural language and code. This training allows UniXCoder to align vulnerability-related text artifacts (like commit messages) with their corresponding code more effectively than models focused on either code or text independently. The model's highest performance is in CP to code mapping (81.72% in top-5), demonstrating its ability to capture complex causal relationships between vulnerability descriptions and code. Its strong performance in AF to code mapping (77.62% in top-5) can also be attributed to its ability to bridge the semantic gap between after-fix descriptions and the corresponding code. Even though its VT to code performance (72.57% in top-5) is slightly lower, UniXCoder maintains consistent results across all entity types.

Table 5. Average VulnExtract and VulnTrace combined experimental result for VT_Code, AF_Code and CP_Code.

| Approach | avg. VT_Code % | | | | | avg. AF_Code % | | | | | avg. CP_Code % | | | | |
|---|---|---|---|---|---|---|---|---|---|---|---|---|---|---|---|
| | top1 | top2 | top3 | top4 | top5 | top1 | top2 | top3 | top4 | top5 | top1 | top2 | top3 | top4 | top5 |
| UniXCoder | **10.82** | 14.54 | **28.71** | 46.81 | **64.0** | **14.36** | **29.93** | **42.32** | **53.68** | **66.74** | 10.14 | 24.53 | **44.46** | **56.71** | **72.03** |
| CoCoSoDa | **10.82** | 13.07 | 20.67 | 30.40 | 43.42 | 7.56 | 17.18 | 27.43 | 37.50 | 48.01 | 7.93 | 19.36 | 27.43 | 38.71 | 50.87 |
| GraphCodeBERT | 6.24 | **14.68** | 28.48 | 33.64 | 58.97 | 2.71 | 8.95 | 20.95 | 33.19 | 44.46 | 8.56 | 17.45 | 37.84 | 52.41 | 63.05 |
| CodeBERT | 2.31 | 7.76 | 27.97 | **50.76** | 59.09 | 4.75 | 11.90 | 23.95 | 38.92 | 52.76 | 13.57 | 24.54 | 37.78 | 51.14 | 67.95 |
| RoBERTa | 4.05 | 9.38 | 26.93 | 46.62 | 53.73 | 5.45 | 14.34 | 25.07 | 36.71 | 53.28 | 11.05 | 22.92 | 36.20 | 51.98 | 65.97 |
| RoBERTa-Code | 9.08 | 13.03 | 20.02 | 35.78 | 61.75 | 5.25 | 11.34 | 23.65 | 36.21 | 51.22 | **15.55** | **26.07** | 39.53 | 51.67 | 66.82 |

Table 6. Average VulnExtract and VulnTrace combined experimental results for VT/CP_Code and AF/CP_ Code.

| Approach | avg. VT/CP_Code | | | | | avg. AF/CP_Code | | | | |
|---|---|---|---|---|---|---|---|---|---|---|
| | top1 | top2 | top3 | top4 | top5 | top1 | top2 | top3 | top4 | top5 |
| UniXCoder | **4.66%** | 6.88% | 19.54% | 29.99% | **59.60%** | **1.99%** | **9.66%** | **21.62%** | **33.56%** | **53.10%** |
| CoCoSoDa | 2.27% | 5.36% | 14.67% | 25.90% | 42.27% | 1.08% | 5.02% | 11.04% | 22.46% | 33.00% |
| GraphCodeBERT | 0.76% | **9.49%** | **27.01%** | 30.53% | 57.28% | 0.42% | 2.98% | 12.35% | 21.72% | 32.83% |
| CodeBERT | 0.0% | 3.97% | 14.82% | **36.31%** | 52.32% | 1.08% | 3.87% | 14.23% | 28.11% | 41.27% |
| RoBERTa | 1.95% | 3.69% | 19.85% | 31.50% | 43.82% | 0.84% | 7.19% | 18.51% | 26.09% | 39.19% |
| RoBERTa-Code | 2.33% | 4.93% | 9.0% | 29.05% | 57.45% | 1.75% | 4.11% | 10.61% | 25.77% | 39.57% |

In comparison, **CoCoSoDa** leverages contrastive learning to differentiate between similar code snippets, which helps in AF and CP mapping but falls short in VT to code (48.51%), likely because it is not optimized for mapping vulnerability triggers effectively. **CodeBERT**, though designed for both code and text, lacks the fine-grained cross-modal training of UniXCoder, leading to lower performance, particularly in AF to code (59.69%). While **RoBERTa** is a robust pre-trained language model, it underperforms in code-specific tasks due to its focus on natural language rather than programming languages. On the other hand, **RoBERTa-Code**, which has undergone code-specific fine-tuning, shows improvements in CP to code (73.05% in top-5), but still trails behind UniXCoder. This gap arises because CodexGlue, while adapted for code, does not fully bridge the gap between vulnerability-specific text and code mapping as effectively as UniXCoder's cross-modal pre-training, which is designed to handle both natural language and code more seamlessly.

Table 4 presents the performance of pairwise sentence-to-code pair mapping for VT/CP_Code and AF/CP_Code. Given that the performance of these pairwise mappings is inherently tied to the individual VT, AF, and CP to Code mappings (as outlined in the Evaluation Metrics), it is expected that UniXCoder, with its superior results in single-entity mappings, outperforms all other code search models in top5 accuracy, reaching over 68.20% in top 5 for both VT/CP_Code and AF/CP_Code.

### 7.3 Performance of VuleExtract+VulnTrace (RQ4)

For our evaluation of RQ4, we prioritized recall in selecting the most effective approach for VulnExtract, as high recall is crucial when dealing with software vulnerabilities. Ensuring that VulnExtract identifies as many relevant entities as possible is key to understanding and addressing potential security risks [41, 52, 54]. Based on the results in Table 2, we opted to use the SVM+patterns VulnExtract output for AF and CP, which yielded strong recall scores of 91.03% and 95.01%, respectively. For VT, we chose the heuristic VulnExtract approach, as it achieved a perfect recall of 100%. These choices reflect our focus on maximizing the identification of vulnerability-related entities, which directly impacts the overall effectiveness of the VulnTrace system.

As shown in Table 5, UniXCoder outperformed the other models across VT_Code, AF_Code, and CP_Code in most top-k categories. In particular, UniXCoder achieved 14.36% accuracy for top-1 AF_Code matching and 66.74% for top-5 AF_Code matching, indicating that it is highly effective at retrieving the correct code related to AF vulnerabilities. Its strong performance continues with top-5 CP_Code retrieval, reaching 72.03% and top-5 VT_Code reaching 64.0%, which highlights its ability to identify critical code for AF, CP and VT entities. RoBERTa-Code also showed competitive results, particularly for VT_Code achieving 61.75% for top-5. However, it falls behind UniXCoder in AF_Code and CP_Code retrieval, indicating that while they can be effective in certain contexts (e.g., VT_Code), their overall utility in identifying vulnerability-related code may be more limited when AF and CP entities are considered.

Table 6 highlights the combined performance for VT/CP_Code and AF/CP_Code. UniXCoder once again stands out with the highest top-5 scores for both VT/CP_Code (59.60%) and AF/CP_Code (53.10%). These results indicate that UniXCoder is the most suitable pre-trained model for the code search tasks in VulnTrace. Trained on both natural language and programming languages using a cross-modal pre-training approach, UniXCoder leverages its understanding of code semantics and structure, making it particularly effective for identifying vulnerability-related code snippets in static analysis. Furthermore, these results demonstrate that the combination of VulnExtract and VulnTrace is highly effective in identifying and tracing a wide range of vulnerabilities. By leveraging VulnExtract's high recall for extracting pairs of vulnerability-related entities and VulnTrace's ability to map those entities to relevant code lines, this approach offers a comprehensive and reliable solution for static vulnerability tracing in complex software systems.

*Implication.* We designed a controlled experiment to evaluate the impact of our tool on vulnerability comprehension while ensuring fairness among participants. Six Computer Science Ph.D. students participated in the study, completing the task under two conditions: without tool support **(Group A)** and with tool support **(Group B)**. To ensure fairness and minimize the influence of prior knowledge, each participant performed the task with different CVE sets in the two conditions, allowing for a balanced comparison. A total of 40 CVEs were selected for the study—six randomly chosen from each project and four from the full dataset. The 40 CVEs used in the study were divided into six parts, with the first two parts containing 10 CVEs each and the remaining four parts containing 5 CVEs each. Each student was assigned one part in Group A and a different part in Group B, ensuring that no participant analyzed the same CVEs in both conditions. This setup allowed all participants to experience both experimental conditions while avoiding repeated exposure to the same vulnerabilities. Each participant was provided with vulnerability-related artifacts, including the CVE summary, bug report, commit message, and code snippets. Their task was to identify VT (Vulnerability Trigger), CP (Crash Phenomenon), and AF (After-Fix) for each CVE and map them to the corresponding code lines. Afterward, they were required to write a Fix Summary that integrated VT, CP, and AF with their mapped code lines. The procedure followed by each group is described below:

- Group A (Baseline, without tool support):
    - Manually examined all artifacts.
    - Carefully read and analyzed all available information.
    - Identified VT, CP, and AF along with their corresponding mapped code lines.
    - Composed the Fix Summary based on their manual understanding.
- Group B (With tool support):
    - Reviewed artifacts but skipped lengthy sections when applicable.
    - Relied on VT, CP, and AF sentences extracted by VulnExtract to form an initial understanding.
    - Used top 5 mapped code statements from VulnTrace to locate relevant code efficiently.
    - Cross-verified extracted information with the artifacts and finalized their Fix Summary.

Each student recorded the total time spent per CVE, including artifact analysis and summary writing. Two authors independently evaluated their performance based on:

- Correct identification of VT, CP, and AF with mapped code lines (3 points).
- Completeness and clarity of the Fix Summary (2 points).

The final score was determined by averaging the two authors' ratings. The results demonstrate that our tool could improve both efficiency and accuracy in vulnerability comprehension. Participants in Group A spent an average of 7.09 minutes per CVE and achieved an average score of 3.46. In contrast, Group B, aided by our tool, reduced analysis time to 3.44 minutes while improving their performance score to 3.88. These findings suggest that our tool may help users navigate extensive NL artifacts more efficiently, allowing them to focus on key vulnerability-relevant information. By localizing bug-fix-related text and linking it to corresponding code, the tool reduces manual effort and improves comprehension.

## 8 THREATS TO VALIDITY

A primary threat to *construct validity* lies in the subjectivity involved in discourse pattern extraction and the labeling of vulnerability entities (AF, VT, CP). To mitigate this, we implemented a rigorous annotation process (Section 4.1), where two independent coders labeled each sentence. We assessed coding reliability using inter-rater agreement metrics, achieving Cohen's Kappa values of 0.83, 0.84, and 0.84 for AF, VT, and CP, respectively, as detailed in (Section 4.1.3). Disagreements were reviewed collaboratively to ensure consensus and minimize subjectivity. Regarding discourse pattern extraction, our process adhered to open coding practices [10], which emphasize iterative analysis and refinement to reduce coder bias and improve consistency.

*Internal validity* threats primarily concern the implementation of VulnTrace and the potential influence of the varied performance of different code search models. To address this, we evaluated six state-of-the-art pre-trained models (Section 5.2.1). By using multiple models and reporting average results, we mitigated the risk of overfitting or bias introduced by any specific model.

*External validity* pertains to the generalizability of our findings. Although our dataset includes vulnerabilities extracted from existing CVEs, these vulnerabilities may not comprehensively represent all possible software vulnerabilities. To enhance generalizability, we adopted an LOOCV strategy across six diverse open-source datasets. This approach helped ensure that identified patterns generalize well across projects and reduced the risk of overfitting to any single dataset. Nevertheless, we acknowledge this limitation and suggest that future studies expand the dataset to include vulnerabilities discovered through additional methods.

## 9 RELATED WORK

**Analyzing vulnerabilities in natural language artifacts:** Several studies have focused on analyzing natural language artifacts such as bug reports, CVE summaries, and commit messages to identify bug-fix patterns and improve the bug resolution process [5, 22, 39, 44, 64, 67]. One of the foundational works investigates the key attributes of high-quality bug reports and their relation to the resolution of software vulnerabilities. It highlights how well-written bug reports can lead to faster and more effective fixes [5]. Similarly, the study examines how natural language discussions in bug reports contribute valuable context to the bug-fixing process [44]. An empirical study uses natural language processing (NLP) techniques to analyze commit messages and bug reports, automatically discovering vulnerabilities based on patterns identified in the textual descriptions [67]. Gousios et al. explores how emotional variations in commit messages can provide early indicators of potential vulnerabilities [22]. These studies underscore the importance of leveraging natural language artifacts in understanding and resolving software vulnerabilities. Our approach, *VulnExtract*, automates the extraction of key vulnerability entities — Vulnerability-Trigger (VT), Crash-Phenomenon (CP) and After-Fix (AF), from such artifacts, while capturing the relationships between a pair of entities.

**Analyzing vulnerabilities in code or extracting code patterns:** Code analysis is crucial in identifying patterns associated with vulnerabilities and their corresponding fixes [7–9, 34, 40, 42, 49]. Nguyen et al. present a large-scale empirical study of common bug-fix patterns, identifying patterns that frequently resolve vulnerabilities in code [8]. Zhong et al. compare vulnerability-specific patches with general bug fixes, showing that security patches often involve unique code patterns [7].

Furthermore, Patchworking highlights how patterns in code changes can be generalized across different vulnerabilities, providing insights into recurring fix behaviors [9]. These works highlight the need for understanding code patterns to effectively address and fix security vulnerabilities [7–9, 34, 40, 42, 49]. Our approach, *VulnTrace*, automatically traces a pair of related vulnerability entities (VT, CP, AF) in natural language artifacts to their corresponding code segments to recover the semantic relationship between lines of code.

**Traditional traceability models from natural language artifacts and code:** Traditional traceability models primarily establish/recover tracing links between individual entities, such as between requirements or bug reports and corresponding code segments [5, 11, 46, 61]. Bettenburg el at. explores how bug reports can be linked to their corresponding code fixes, laying the foundation for traceability models that focus on one-to-one mappings [5]. VCMatch proposes a method for linking security patches to their corresponding vulnerabilities, laying the foundation for traceability models focusing on one-to-one mappings between artifacts and code [61].

While these studies have advanced our understanding of traceability models, they are limited to single-entity-to-code mappings [5, 11, 46, 61]. In contrast, our approach extends these traditional models by focusing on pairs of entity relationships, such as VT/CP and AF/CP, and mapping them to corresponding related lines of code. This enhances the ability to track and understand bug fixes more efficiently, expanding the scope of traceability models in the context of security vulnerabilities.

## 10 CONCLUSIONS AND FUTURE WORK

This paper introduces a traceability model that links a pair of sentences describing VT, CP and AF to their corresponding pair of code statements facilitating a deeper understanding of how and why vulnerabilities are fixed. Our empirical study, based on 341 CVEs and their associated code snippets, demonstrates the effectiveness of our approach, with recall exceeding 90% in most cases for NL sentence extraction. The end-to-end combined VulnExtract+VulnTrace achieves a Top5 accuracy of 59.6% and 53.1% for mapping two pairs of NL sentences to code statements. Future work will explore further optimization of the VulnTrace model, focusing on improving its handling of complex and abstract NL vulnerability descriptions such as VT. Additionally, we aim to extend the model's applicability to other domains and software systems, enhancing its robustness and scalability for large-scale real-world applications.

## DATA AVAILABILITY

The replication package of this work and additional experiment results are available at [16]. We will make the implementation and data publicly available upon acceptance.